# Carrington's Last Sunspot Drawing in 1870

Hisashi Hayakawa[1] (1-4), E. Thomas H. Teague[2] (5)

(1) Institute for Space-Earth Environmental Research, Nagoya University, Nagoya, Japan

(2) Institute for Advanced Research, Nagoya University, Nagoya, Japan

(3) Space Physics and Operations Division, RAL Space, Science and Technology Facilities Council, Rutherford Appleton Laboratory, Harwell Oxford, Didcot, Oxfordshire, UK

(4) Nishina Centre, Riken, Wako, Japan

(5) WDC-SILSO[3] observer, Chester, UK

**Abstract**

Richard Christopher Carrington carried out extensive sunspot observations from 1853 to 1861 and left numerous scientific legacies. However, despite his fame, consulting his original manuscripts, we have found Carrington's last sunspot drawing, which he never published. We date this drawing 23 Dec 1870, when Carrington was based at his Churt Observatory. The drawing allows us to count 3 sunspot groups and 10 individual sunspots and measure the group positions at N14° W37°, N11° W46°, and S08° W08°, where our longitude values represent longitudinal distance from the central meridian. In comparison with the existing datasets, his raw group counts look slightly lower than the contemporaneous observers. His position data look consistent. These additions contribute to the ongoing efforts for the sunspot number recalibration and butterfly diagram constructions, as long as we follow the standard assumption for the homogeneity of the "Schmidt backbone". Otherwise, this case instead questions the validity of the common assumption of "Schmidt backbone" keeping the data homogeneity. This sunspot drawing is unlikely to represent the tip of a huge, lost iceberg, but rather an isolated drawing that he left from his Churt era. Carrington may have temporarily revived his interest upon the solar eclipse of 22 Dec 1870. Otherwise, Carrington's references to his sunspot research in his Churt era are confined to priority claims in response to Chevalier's RAS obituary and a local newspaper report. The former confirms that Carrington was probably the first person, who counted and published sunspot data from Thomas Harriot's original sunspot drawings.

[1] https://orcid.org/0000-0001-5370-3365
[2] https://orcid.org/0009-0002-4304-425X
[3] Here we use "SILSO" to abbreviate "Sunspot Index and Long-term Solar Observations".

## 1. Introduction

Richard Carrington (26 May 1826 – 27 Nov 1875) was, without doubt, one of the giants not only of Victorian astronomy but also of solar physics. The extent to which modern solar physics lives benefits from Carrington's legacy has been reviewed and highlighted in several recent publications (Cliver, 2006; Clark, 2009; Cliver and Keer, 2012; Seehan, 2014; Cliver et al., 2022). While living at Redhill, Carrington not only compiled a star catalog for the circumpolar stars down to the 10th magnitude (Carrington, 1857) but also monitored sunspots on the solar surface from 1853 to 1861 (Carrington, 1863).

Throughout his observations, Carrington carefully monitored the sunspot positions. Carrington was responsible for conceptualising the equatorward migrations of the sunspot belts in each solar cycle (Carrington, 1858, 1859b), currently known as Spörer's law and a key marker for the temporal evolution of solar cycles (Hathaway, 2015). Carrington also monitored the longitudinal motion of the sunspot groups. Following his observations, solar rotations are now referred to as "Carrington rotations". His observations also disclosed evidence of progressively slower rotation as heliographic latitude increases from the equator towards the poles (Carrington, 1859b, 1862), a variation now known as "differential rotation" (Beck, 2000).

Carrington was also one of two serendipitous witnesses of the earliest observed solar flare (Carrington, 1859a; Hodgson, 1859). This flare has been named the "Carrington Flare" in his honour. It was one of the most powerful solar flares in observational history, bringing one of the greatest geomagnetic storms ever recorded. This flare still serves as a benchmark for analyses and discussions of extreme solar storms (Tsurutani et al., 2003; Cliver and Svalgaard, 2004; Cliver and Dietrich, 2013; Curto et al., 2016; Hudson, 2021; Hayakawa et al., 2022, 2023; Cliver et al., 2022; Usoskin et al., 2023; Hudson et al., 2024, 2025).

We still have much to learn from Carrington's records and achievements. Modern scientists have frequently consulted Carrington's publications (Carrington, 1859a, 1863). Carrington's sunspot position data have been digitised (Casas and Vaquero, 2014) and contributed to reconstruction of centennial solar-cycle variations (Muñoz-Jaramillo and Vaquero, 2019). Thomas Teague documented his measurement methods (Teague, 1996) and recounted Carrington's sunspot drawings (1853-1861) from Carrington (1863)[4] to publish his recounts in Bhattacharya et al. (2021), contributing to recalibration of the relative sunspot number (Clette et al., 2023). In addition, Carrington's archival records allow us to analyse the Carrington flare in detail (Hayakawa et al., 2018; Prosser, 2018), to measure the source location as N21° W14° and the sunspot group area as 2971 millionths of the solar hemisphere, and to estimate the flare's magnitude as X80, ranging between X46 and X126 (Hayakawa et al., 2023a).

Until recently, despite Carrington's accomplishments, no one knew what he looked like. As pointed out by Cliver et al. (2021), a photograph of Lord Kelvin was mistakenly taken to be that of Carrington[5]. It was not until 2026 that the only known photograph of Carrington (Figure 1) finally came to light (Bond and Bowler, 2026). This is also the case with his sunspot drawings. What Teague recounted were the sunspot observations Carrington reproduced in his book (Carrington, 1863; Bhattacharya et al., 2021). His whole-disk sunspot drawings have remained in the Royal Astronomical Society Archives and are in want of further investigation. In fact, Carrington did not publish the entirety of his sunspot observations in his book. In addition, we have located an overlooked sunspot drawing from 1870 that Carrington did not publish. The present study documents this late sunspot drawing (Section 2) and his observational background at that time (Section 3). This study also measures the relative sunspot number and sunspot group positions from his drawing (Section 4), contextualises his observation to modern reconstructions of the contemporaneous solar activity (Section 5), reviews his sunspot-related works during his Churt era

[4] By reference to Carrington's published drawings and measurements (Carrington, 1863), Teague has estimated the sunspot number according to the formula 10g + f for each hemisphere on each day of observation. Subject to necessary omissions (such as the quality of the 'seeing', which Carrington did not record), Teague applied the standard procedure as currently used by members of the WDC-SILSO observing network. Wherever available, Teague relied upon Carrington's daily drawing sequences in Carrington (1863). For groups observed only once, he used the relevant rotation drawings. In the very rare cases where no drawing of either kind was available (usually groups extremely close to the solar limb), he referred to Carrington's measured positions together with any additional notes. Carrington's original sunspot drawings in the RAS archives are yet to be studied systematically.

[5] See e.g., https://www.solarstorms.org/SCarrington.html

(Section 6), and hypothesises why he left this sunspot drawing specifically in 1870 (Section 7).

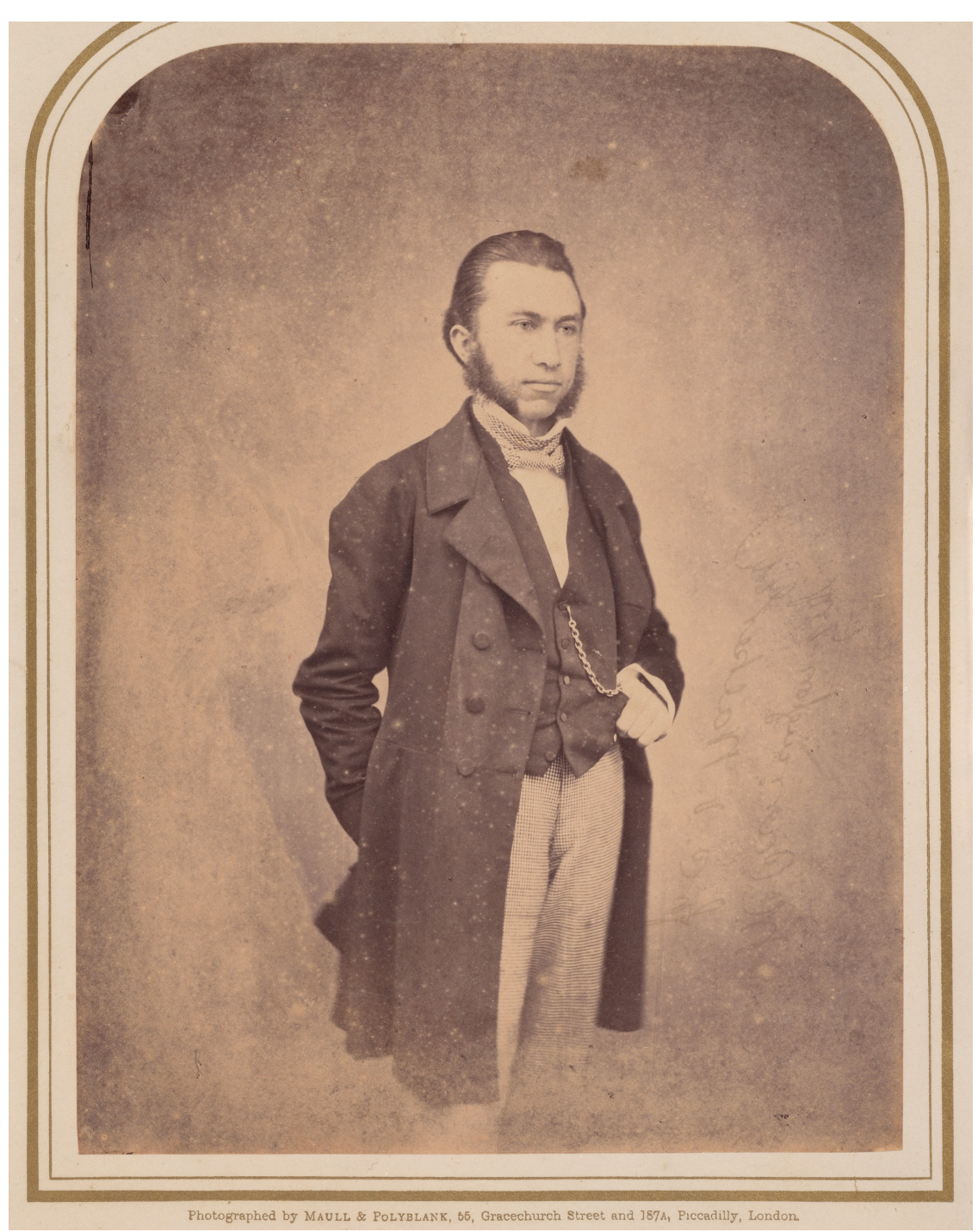

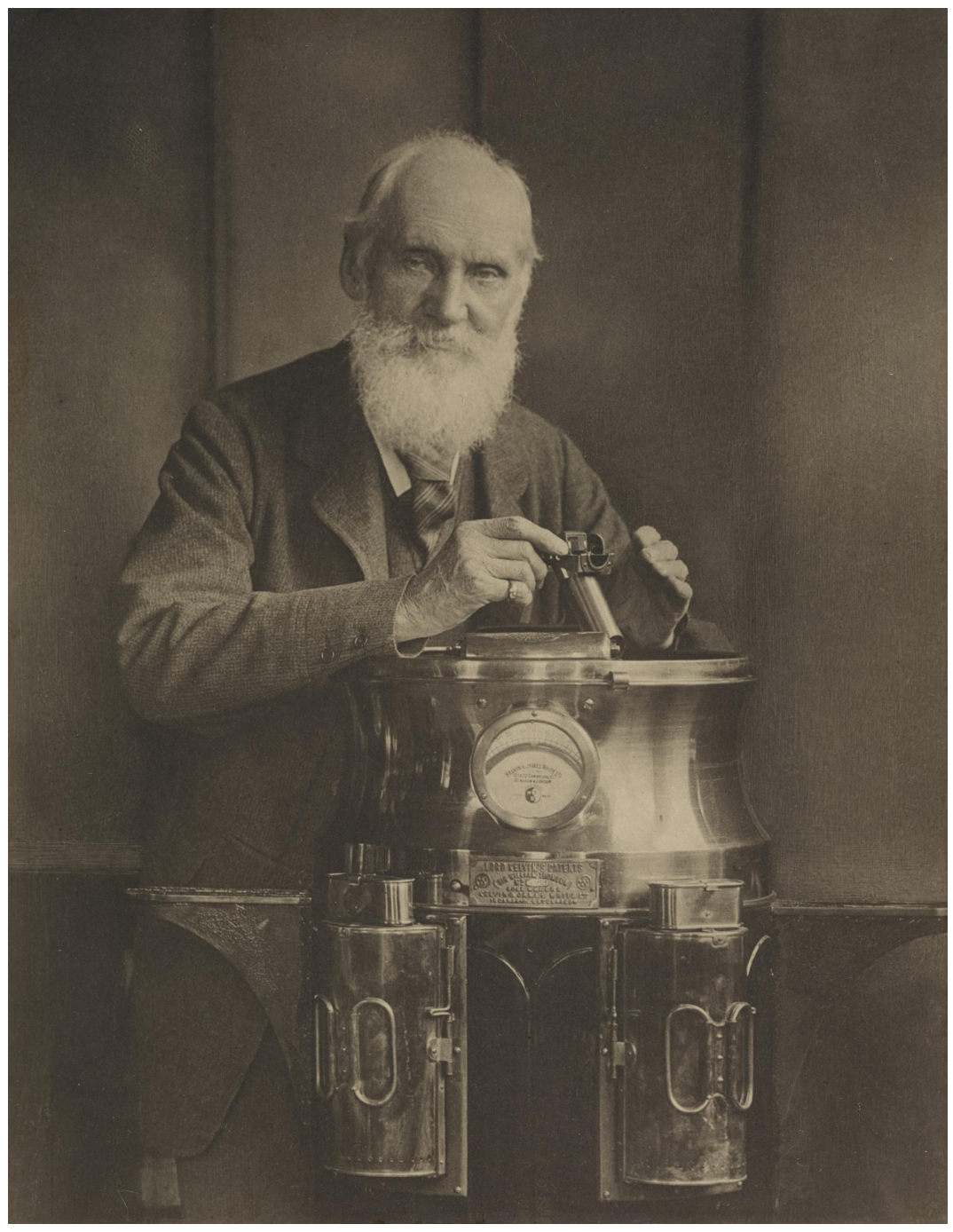

Figure 1: Richard Carrington's photo (left: MSS RAS) and a photograph of Lord Kelvin that was misidentified as Carrington in some publications (right: National Galleries of Scotland, PGP 230.1; reproduced by courtesy of the Royal Astronomical Society and the National Galleries of Scotland).

## 2. Carrington’s Sunspot Drawing on 23 Dec 1870

Figure 2 shows Carrington’s last sunspot drawing. We found it in the Royal Astronomical Society Archives and time and date it at 12:16 LAT (local apparent time) on 23 Dec 1870. This drawing was made after Carrington left Redhill Observatory (Carrington, 1863). His Redhill solar observations (Figure 3) extended from 9 Nov 1853 to 24 Mar 1861 (Carrington, 1863). That is what Carrington published and is what the scientific community has understood (Teague, 1996; Vaquero et al., 2016; Arlt and Vaquero, 2020; Bhattacharya et al., 2021; Clette et al., 2023). However, not everything has in fact been published. Carrington’s original records were sent to and preserved in the Royal Astronomical Society Archives (ACR, 1876, p. 250). Here, Carrington’s sunspot records are sorted into four categories, as reproduced verbatim from Bennett (1978, p. 33):

1. Sunspot observations, 3 vols, 1853–61.

2. Reductions of these observations, 7 vols, 1853–61 (with one addition, 1873).

3. Drawings of sunspots, showing the whole of the Sun's disk, 3 folio vols, 1853–61, and one 1870.

4. Drawings of groups of sunspots, folio vol.

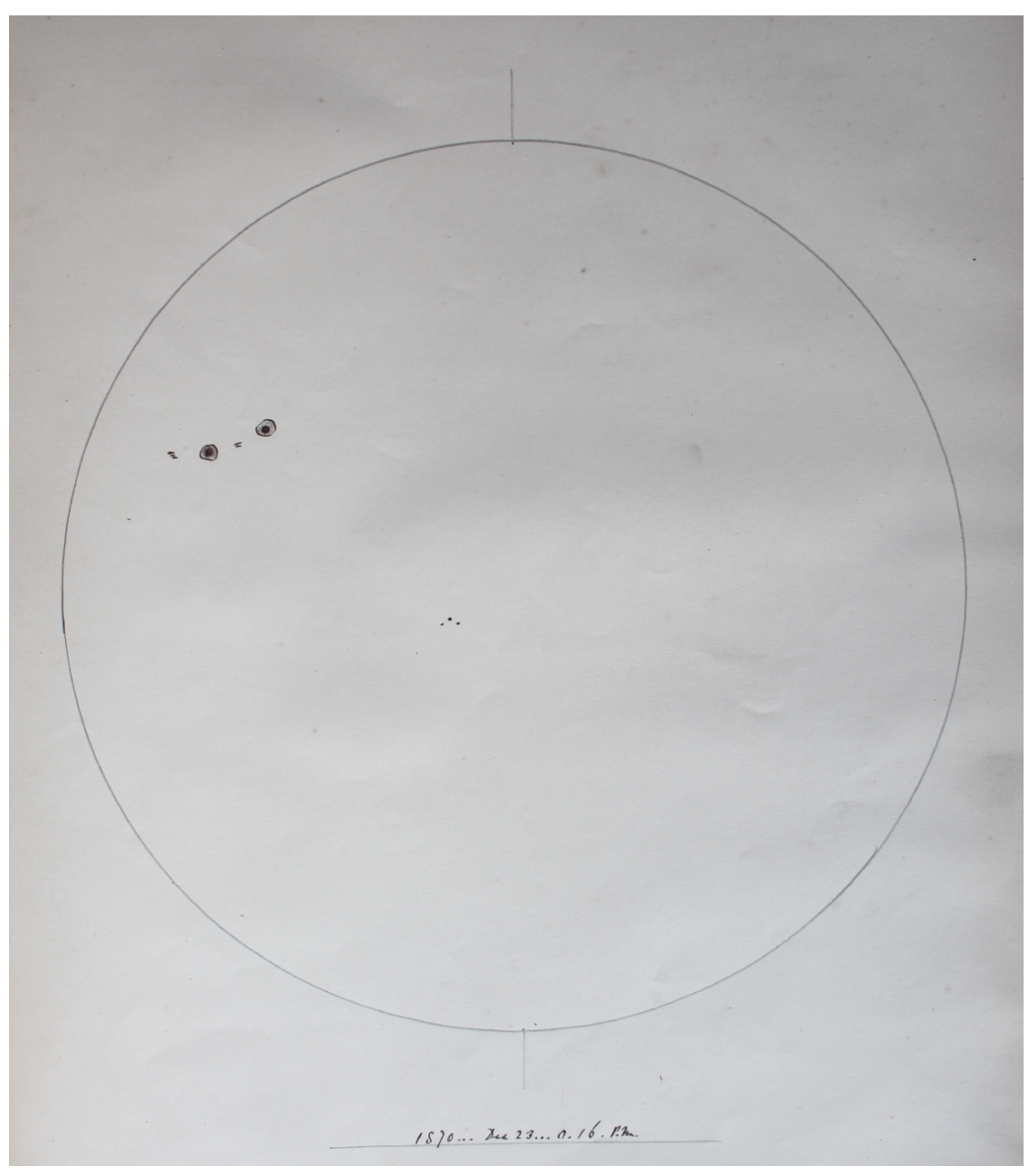

Figure 2: Carrington's last sunspot drawing on 23 Dec 1870 (MSS Carrington 3.3, f. 277a), reproduced by courtesy of the Royal Astronomical Society.

Among them, MSS Carrington 1 and MSS Carrington 3 are of significant interest for quantification of the sunspot parameters, as shown in Figure 1 of Hayakawa et al. (2023). MSS Carrington 1 and MSS Carrington 3 comprise his notebooks for the individual sunspot groups and his whole-disk sunspot drawings, respectively. This catalog mentions one whole-disk sunspot drawing in 1870. It is MSS Carrington 3.3 (f. 277a) where we found Carrington's isolated sunspot drawing in 1870 (Figure 2).

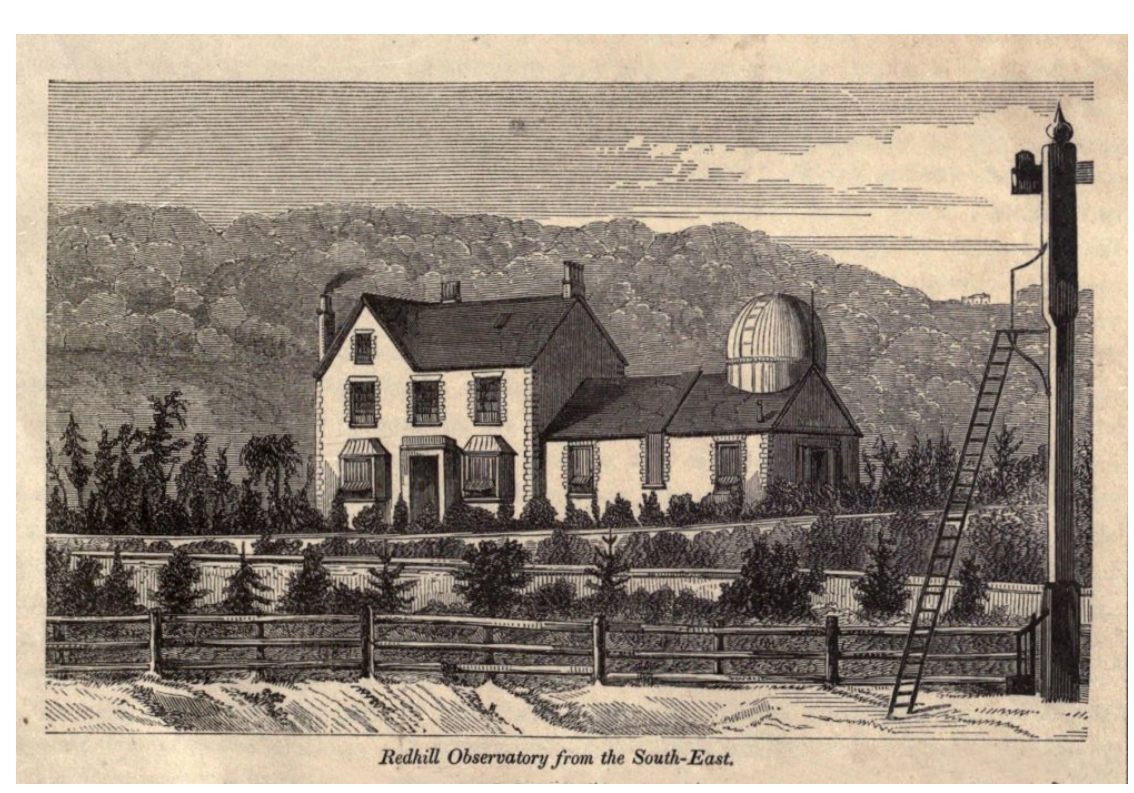

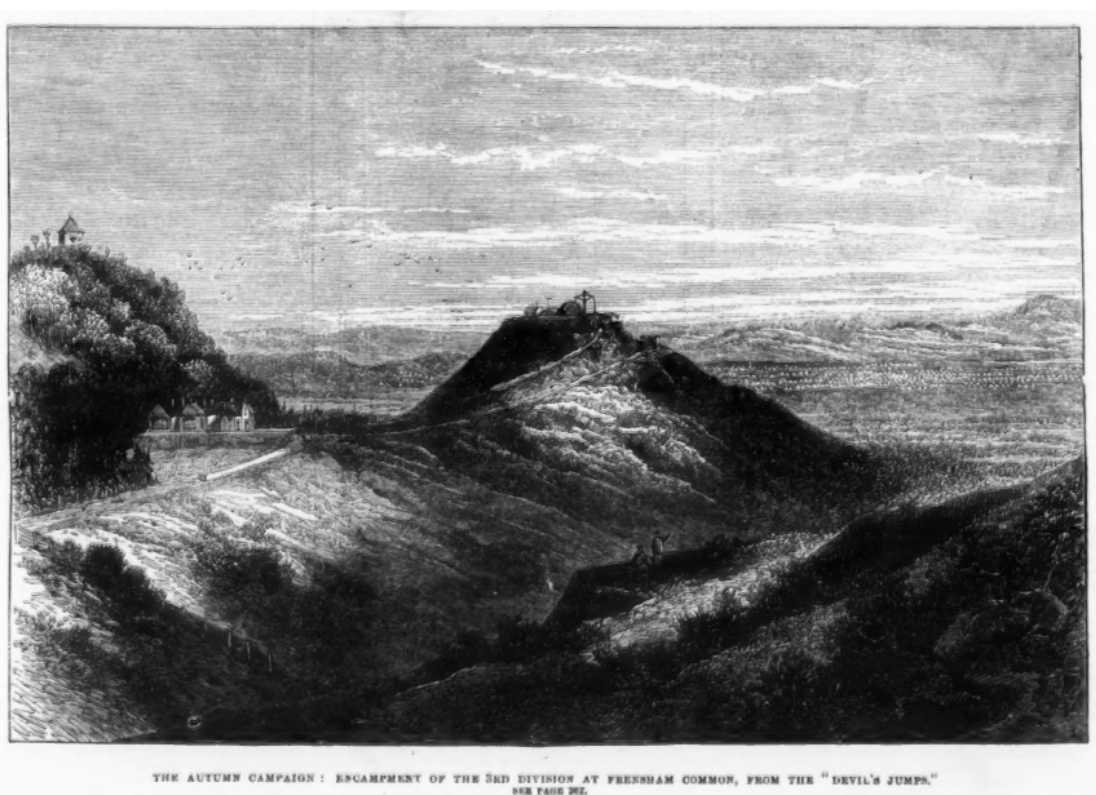

Figure 3: Carrington's Observatories at Redhill and Churt, as publicised in the preface of Carrington (1857) and an illustration of ILN (1871).

**3. Observational Background**

This finding is slightly surprising, as Carrington had shut down his observatory, having quit astronomy and sold his house at Redhill (Figure 3) on 14 Jul 1861 (Clark, 2009, pp. 115-121; Cliver and Keer, 2012, p. 15). The house at Redhill was known as the "Dome House" of Furze Hill, at least in the local topographical narrative (Philipps, 1885, p. 73), and was later replaced by an apartment compex called "the Dome"[6] (Moore, 1999, p. 69). We located the site of Carrington's Redhill Observatory at N51°14'29" W000°10'21" in the World Geodetic System 84. Carrington (1861, p. 3) himself describes the end of his observatory as follows: "On his departure [Sc. Schroeder, his last assistant in Redhill] in March 1861, which occurred sooner than expected, in consequence of his being recalled to become Director of a provincial school, I decided to close the series and wind up the results I had obtained" (Carrington, 1863, p. 3). The telescopes that Carrington had used to measure star positions ended up in the Radcliffe Observatory before moving to the Museum of the History of Science at Oxford (Hutchins, 2008, pp. 519-520; Clark, 2009, p. 115).

Where was Carrington based in 1870, when he made his last sunspot drawing? In 1869, after shutting down the observatory at Redhill, Carrington started work on establishing a new one (Figure 3) at Middle Devil's Jump near Churt (Carrington, 1869, p. 43). His new observatory was depicted in *Illustrated London News*, as shown in Figure 3. Carrington located it *inside* the hill. His explanation reads as follows: "Being on a hill I did not want elevation, so I have sunk the

[6] https://www.geograph.org.uk/photo/725092

Observatory below ground, just peeping out over the soil. But I have further sunk a dry well, six feet in diameter, to the depth of forty feet from the centre of the Observatory, and with a horizontal shaft communicating with the south side of the hill, 166 feet in length, closed with three doorways. This is principally intended for the clock, for I am determined that one clock at least shall be properly mounted, at a position of invariable temperature and in an air-tight case. I propose to reduce the pressure to twenty-seven inches of mercury" (Carrington, 1869, pp. 43-44). His house is currently known as Jump House and located at N51°08'51", W000°45'49" and the observatory hill is located at N51°08'56", W000°45'45". At the time of writing, his horizontal shaft still existed but was not accessible for safety reasons.

Carrington made limited use of his instruments during his time at Churt. It is difficult to tell whether he used them for solar observations. The only available hints are what Carrington (1869) documented and illustrated (Figure 4). He wrote: "Of the principal instrument I have to announce that I have decided on having an altazimuth, but constructed on a new principle. I have fixed on Steinheil's principle of making the horizontal axis the effective optical axis, by placing the object-glass at one end, and the eye-piece at the other, with a prism outside the object-glass. The casting of the prism took over three months, but Messrs. Chance succeeded at last; six inches aperture is adopted, and six feet three inches focal length. It is some comfort to think that never need the telescope be raised, only turned round, and the observer always under cover. The principal vertical axis is steel, and the bed-plate of cast-iron. The telescope-tube is also cast-iron. The circles are both of thirty inches diameter; the altitude-circle of gun-metal solid throughout, that is without spokes, and cut out from a much larger piece. The microscopes are of five feet focal length, and I hope to try photography with them" (Carrington, 1869, p. 46).

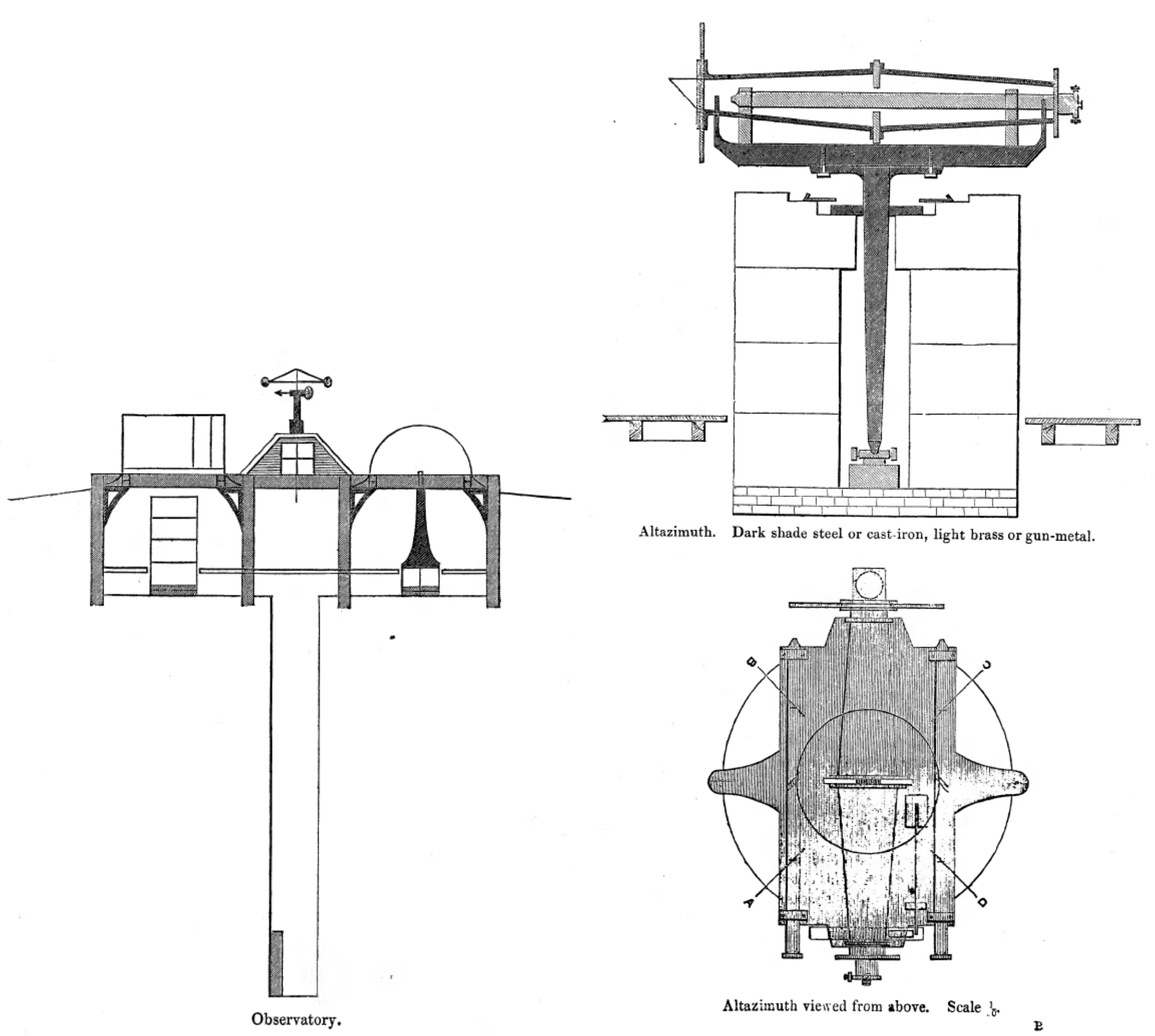

Figure 4: Carrington's own diagrams for his instrumental profiles, as reproduced from Carrington (1869).

**4. Data Extraction**

Carrington's isolated sunspot drawing (Figure 2) is dated 23 Dec 1870. He had made his most recent sunspot drawing before this one nine years earlier, on 23 Mar 1861. Carrington did not publish any results based on the 1870 drawing, on which he recorded the time of observation as 12:16 LAT (local apparent time). From the geographical longitude of Churt Observatory (W000°45'45"), we convert this timestamp in the LMT to 12:19 Universal Time. We counted 3 sunspot groups (G) and 10 individual sunspots (F). The relative sunspot number (R), also known as Wolf Number, is calculated as R = 10G + F (Clette et al., 2023). Applying that rule, we computed Carrington's R on 23 Dec 1870 as 40.

We have derived heliographic coordinates for each of the sunspot groups. Here, the longitudinal values are not in the Carrington longitude system but represent longitudinal distance from the disk center. Figure 2 clearly shows two spikes at the top and bottom of the solar disk. For such measurements, it is always better to minimise geometrical distortion. We followed the guideline of the Royal Astronomical Society Archives and purchased an official digital photo. Unfortunately, this manuscript is too fragile to put on a scanner.

We mirrored Carrington's sunspot drawing, as Carrington's practice was to observe by projection (Carrington, 1863). According to the annotation in 1876, "In all cases, the whole of the Sun's disk has been drawn, and the positions of the north point and the Sun's axis have been marked upon the page with great care" (ACR, 1876, p. 250). In his Redhill era, Carrington projected the solar-disk images, set the celestial north-south axis vertically, and showed the rotation axis and the celestial north-south axes in longer and shorter spikes, respectively, as exemplified in Figure 1a of Hayakawa et al. (2023). We then fitted the solar limb in his drawing to a circle.

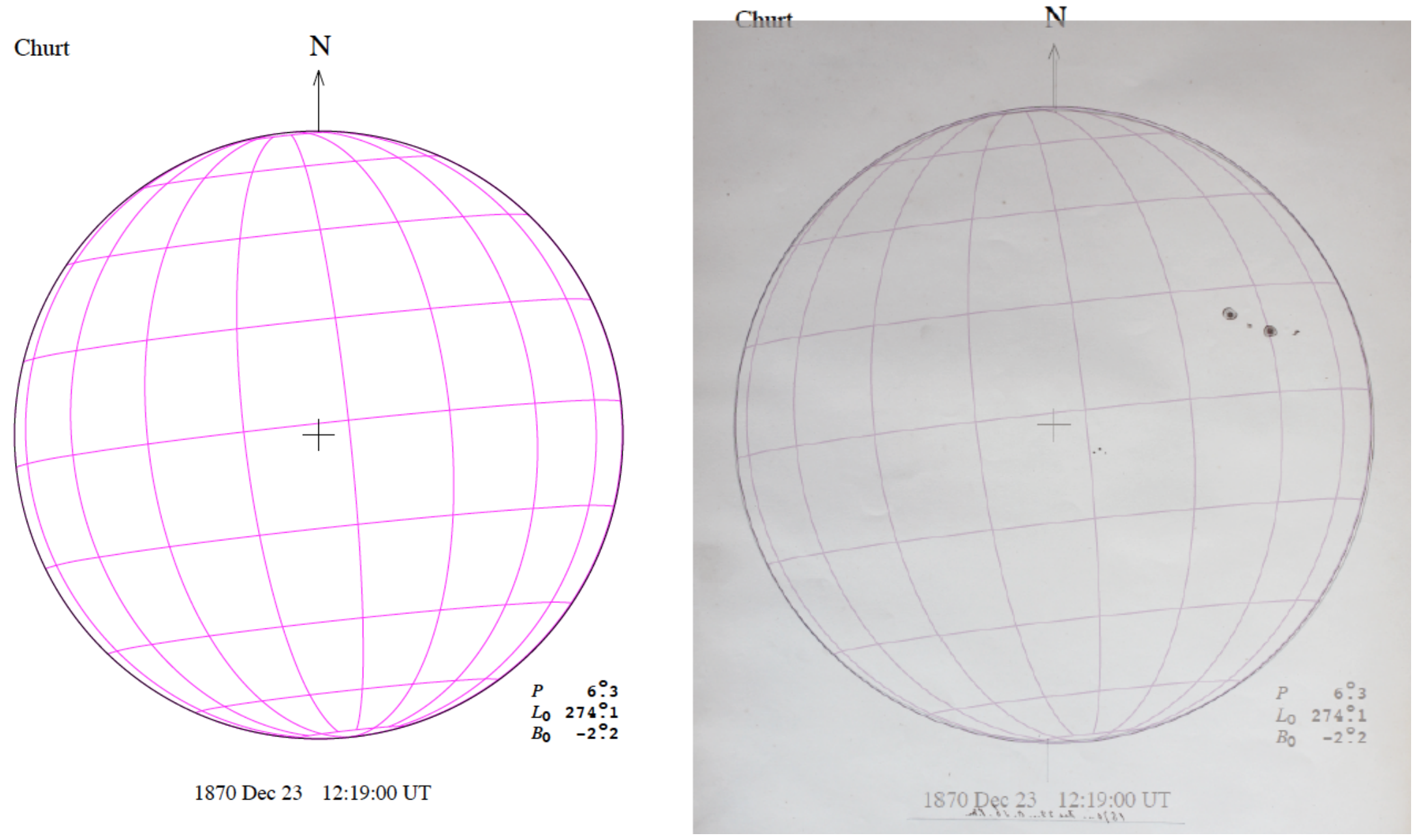

Figure 5: Configuration of the solar disk with the contours of the heliographic latitude and the Carrington longitude as seen from Churt in the local sky at 12:19 UT. The N arrow shows the direction of celestial north. P shows the position angle of the solar north pole. $B_0$ and $L_0$ show values of the latitude and Carrington longitude of the disk centre. On the right panel, Carrington's sunspot

drawing is shown after deprojection and overlaid with the heliographic coordinate grid.

After defining the orientation, we computed the configuration of the solar disk with respect to his sunspot observation (Figure 5a), using the ephemeris data of NASA JPL DE 441 (Park et al., 2021), the Earth's rotation parameter of $\Delta T$ = 2 s (Stephenson et al., 2016; Morrison et al., 2021), the rotation element of Archinal et al. (2011), and the methodology of Hayakawa et al. (2021). We then aligned the sunspot drawing with this plot by reference to celestial north and used it to measure the centroid of each group.

This resulted in the configurations of $B_0$ angle = −2.2° and P angle = 6.3° (Figure 5). We overlaid our calculation (Figure 5) on Carrington's sunspot drawing (Figure 2) and measured the sunspot group positions in this sunspot drawing as N14° W37°, N11° W46°, and S08° W08°, where the longitude values we cite represent longitudinal distance from the central meridian rather than Carrington longitudes.

Even during his time at Redhill, Carrington left no details either of the method by which he plotted groups on his sunspot drawings or of the uncertainty of their positions. Nevertheless, we can attempt a rough estimate. We know that Carrington, who invariably observed by projection, was a skillful and meticulous observer who went to great pains in order to produce drawings that were as accurate as possible (Carrington, 1863). The probability is that the spots shown in his final drawing have been placed within ±1° of their true positions. Conservatively, we can be confident that the uncertainty is unlikely to have exceeded ±2°, as that would represent a significant and avoidable error with spots as small as these when plotted on a disc as large as 12 inches diameter. We could add ±1° of uncertainty for our own measurements on Carrington's sunspot drawing. Therefore, we would expect an uncertainty of ±3° for our results.

## 5. Contextualisation to the Contemporaneous Data

Despite its isolated nature, Carrington's 1870 drawing adds further material to the contemporaneous sunspot data. That year marked the maximum of Solar Cycle 11, where monthly smoothed international sunspot number ver 2 (ISN V2) peaked at 234 in Aug 1870 (Clette and Lefèvre, 2016; Clette et al., 2023). Figure 6 contextualises Carrington's sunspot drawing to the contemporaneous raw group counts in Vaquero et al. (2016) and Teague's recounts. On 23 Dec 1870, only two data have hitherto been known: Schmidt of Athens (G = 8) and Weber of Peckoleh (G = 6), according to Vaquero et al. (2016). Here, Carrington's G looks lower than the average observers in this period. Further investigations are needed.

This data point extends Carrington's data series to 1870 beyond his existing chronological coverage between 1853 and 1861. The extension allows us to revise the timespans of the contemporaneous observers that have been used for the recalibration of ISN V2 from what Figure 1 of Bhattacharya et al. (2021) and Figure 1 of Bhattacharya et al. (2024) show. It also improves Carrington's overlap with Weber of Peckoleh, whose dataset has been used as one of the backbones in Bhattacharya et al. (2024), as explained in their Section 4.3 and Figure 11. Finally, it improves the overlap of Carrington with Schmidt, whose dataset was used as one of the calibration backbones in and Cliver and Ling (2016) and Chatzistergos et al. (2017).

This conclusion is valid — even with the use of a different instrument and observational site from Carrington's main series —, as long as we follow the data quality control of the modern authoritative studies (*e.g.*, Chatzistergos et al., 2017). These studies tend to assume that "differences between the various series are due to different acuity thresholds of the observers" and "the threshold of each observer remains constant throughout the observing period" (Chatzistergos et al., 2017, p. 1), while they also noted this may not be the case. In fact, Chatzistergos et al. (2017) chose Julius Schmidt as one of his backbone observers (see "Schmidt Backbone" in their Figure 1 and Table A.4), while Julius Schmidt moved from Germany to Athens and changed his instruments several times in reality (Schmidt, 1857; Seehan and Dobbins, 2014). Therefore, as long as we follow the data quality control of Chatzistergos et al. (2017), we need to add Carrington's last sunspot drawing — even with uncertainty on his instrumental profile in the Churt era — to Carrington's main series and slightly improve the intercalibration of Carrington data series with other observers' data series, as a matter of consistency.

However, in reality, we are not absolutely certain what telescope Carrington used for his 1870 sunspot observation and the lower group count relative to other observers suggests some loss of visual acuity on his part. Studies on modern long-term sunspot observers also show occasional scale changes upon their changing observational sites and/or instruments (Hayakawa et al., 2020, 2023b). Therefore, Carrington in his last sunspot drawing (Figure 2) may not have maintained the homogeneity with his main sunspot observations in 1853 – 1861 (Carrington, 1863) that Teague recounted. This is contrasted with the validity of the data quality control of the modern authoritative studies (*e.g.*, Chatzistergos et al., 2017). In this case, Carrington's last sunspot drawing might instead question such assumptions, at least those on the "Schmidt Backbone" in such studies, as Julius Schmidt himself moved from Germany to Athens and changed his instruments several times.

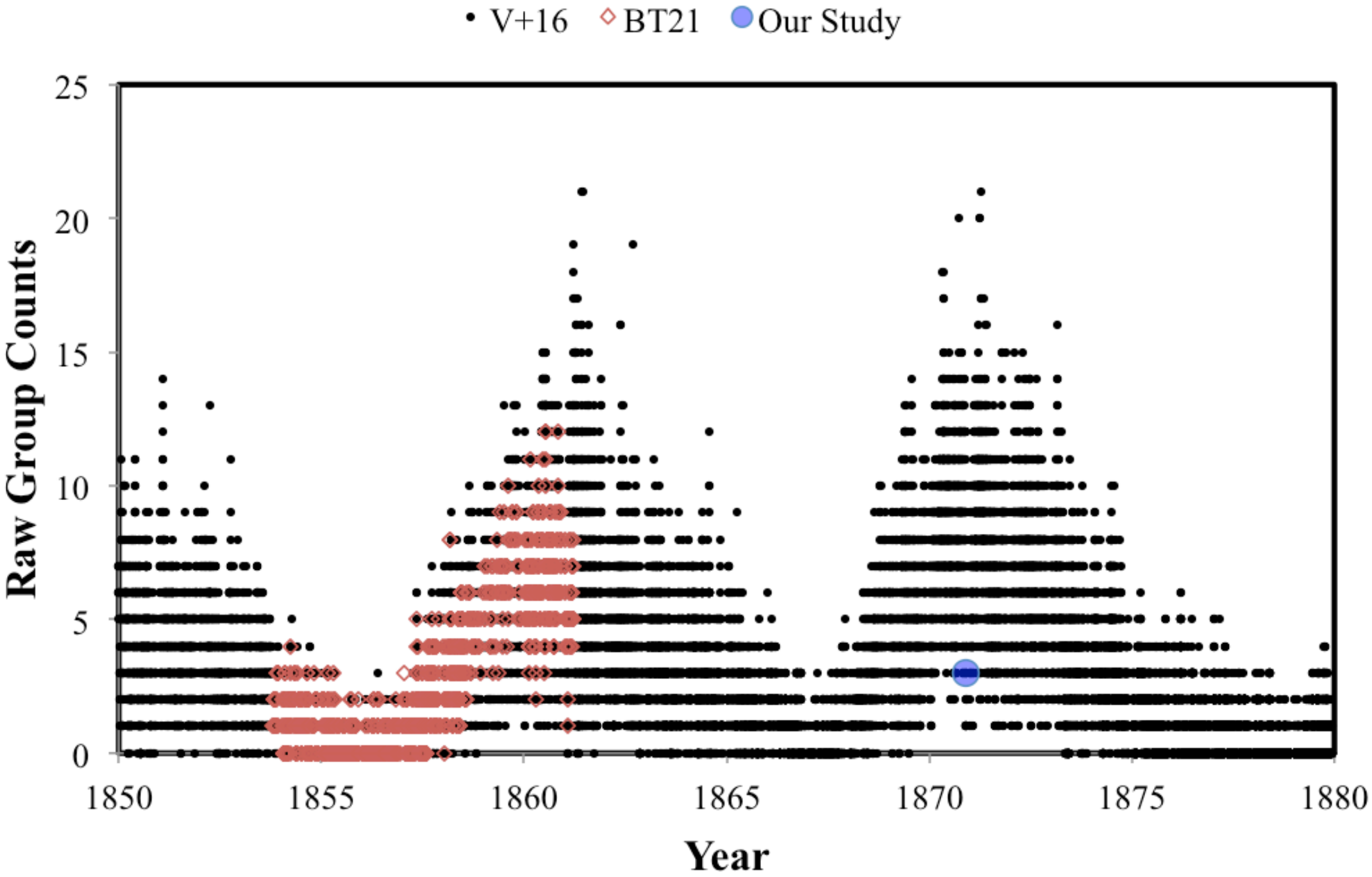

Figure 6: Raw group counts in 1850 – 1879, comparing those of Carrington's last sunspot drawing (a blue circle: ours), Carrington in Teague's recount (red diamonds: BT21), and the known data except for Carrington in V+16 (black dots: Vaquero et al., 2016).

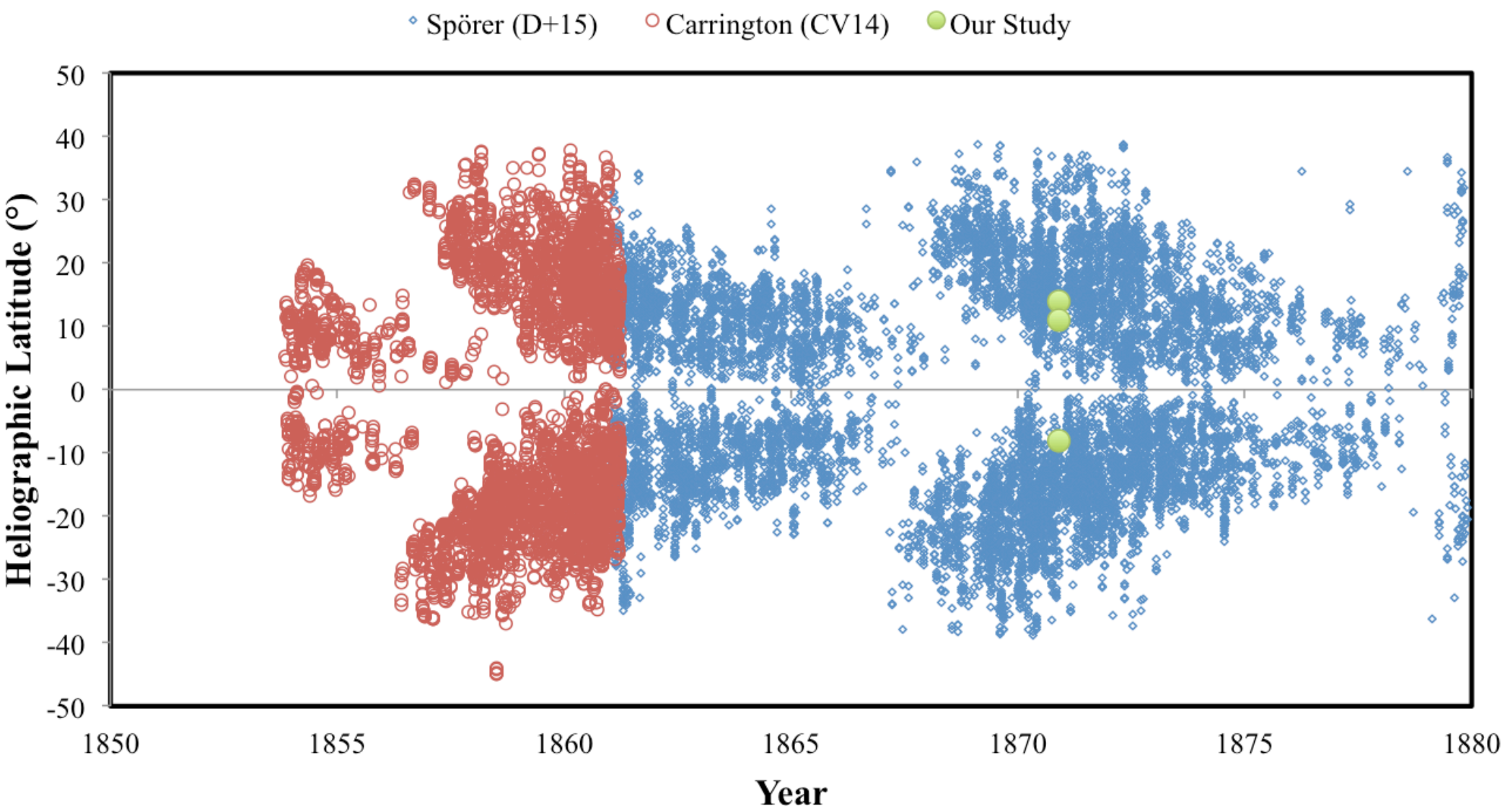

Figure 7: Latitudinal distributions of sunspot group position data of Carrington's last sunspot drawing (a green circle: our study), in comparison with those of Carrington (red circles: Casas and Vaquero, 2014) and Spörer (blue diamonds: Diercke et al., 2015).

Figure 7 contextualises the sunspot group position of Carrington's last sunspot drawing by comparison with contemporaneous data of sunspot group positions. Here, we have acquired the tabulated data of Carrington (1863) from Casas and Vaquero (2014). Among the contemporaneous observers' data, we plotted sunspot group position data of Spörer that span from Jan 1861 to Jan 1894, according to the tabulation of Diercke et al. (2015). In contrast, we have not included de la Rue, Peters, and Schwabe (Casas and Vaquero, 2014; Arlt et al., 2013), as their data series had stopped in Dec 1866, May 1870, and Dec 1867, respectively. This figure locates sunspot groups in Carrington's last sunspot drawing within the sunspot belts at that time – in relatively low-latitude sides in each case – and confirms their consistency with Spörer's contemporaneous sunspot measurements.

## 6. Carrington's Sunspot-Related Works in his Churt Era

This sunspot drawing seems unique and isolated. It is the only sunspot drawing from Carrington's Churt period that we have located in his manuscripts in the Royal Astronomical Society Archives, so far (Section 2). Carrington's published work contains no reference to sunspot observations carried out at Churt, either. He contributed six further articles to the *Monthly Notices of the Royal*

*Astronomical Society* after his report on the Churt Observatory (Carrington, 1869). These concerned instrumentation (Carrington, 1872, 1873a), fogbows (Carrington, 1871), star-position measurements (Carrington, 1873b, 1874b), and his criticism of Chevalier's obituary (Carrington, 1874a).

Of those articles, only one (Carrington, 1874a) had any relevance to sunspots, and this was primarily concerned with the debates about priority arising from Chevalier's obituaries. Against the claim that Chevalier's method had been "adopted by Mr. Carrington (at one time Observer at Durham), who has made a similar series of observations with marked success" (Carrington, 1874a, p. 250), Carrington clarified his own originality and independence, citing his own articles (Carrington, 1854). Here too, Carrington made no reference to any sunspot observations of his own at Churt.

Against the assertion that Chevalier had been "the first to institute in England the regular, continuous observation of the Solar Spots, which has since led to important results" (Carrington, 1874a, p. 250), Carrington cited the case of Thomas Harriot (c. 1560 – 1621), copies of whose records Carrington obtained from Petworth House on 12 Sep 1857 (Carrington, 1874a, p. 250).

In fact, after an initial reference in Hans Moritz von Brühl, von Zach[7], and Rigaud[8] (Von Zach, 1785; Rigaud, 1832), it was Richard Carrington who first extracted data from Harriot's sunspot drawings, visited the Petworth House Archives in Sussex and with the permission of Colonel Wyndham (the owner of Harriot's manuscripts at that time), sent the data to Rudolf Wolf, thus enabling Wolf to establish Harriot as the first telescopic witness of sunspots[9] (Wolf, 1858, pp. 131-135). Harriot's sunspot records have been further highlighted since the late 20th century (*e.g.*, Herr, 1978; Chapman, 1995, 2008).

[7] Von Zach stated as follows: "Yet no one knew or has yet learned that Harriot is an excellent astronomer, indeed, has the distinction of being the first observer in England, and this is what I want to talk about. Most remarkable among his astronomical writings are 199 observations of sunspots with their drawings. Whether Harriot was not the first discoverer of these spots before Galileo and Scheiner, I will not yet decide with certainty; I find every probability so far" (Von Zach, 1785, pp. 153-154); and: "Moreover, it is very likely that Harriot had his telescope from Holland much earlier than Galileo. His very carefully and extensively described observations prove that he had telescopes with magnifications 10, 20, and 30 times greater; at least, no older observations than his exist, and they continue uninterrupted from 8 Dec 1610, to 18 Jan 1613. I have compared the corresponding observations with those of Galileo and found them to be in agreement" (Von Zach, 1785, p. 154-155).

[8] Rigaud (1832, pp. 511-522) analysed Harriot's records for great comets in 1607 and 1618 in detail. Rigaud also referred to Harriot's sunspot drawings: "We should learn nothing new from facsimiles of Harriot's drawings of the solar spots, and of the configurations which Jupiter's satellites presented to him, especially as none of these last are so early as Galileo's" (Rigaud, 1832, p. 522). However, in reality, Harriot's sunspot observations date back to 1610, even before Galileo (Wolf, 1858; Carrington, 1874a; Vaquero et al., 2016). It is not clear whether Rigaud analysed Harriot's sunspot records in detail.

Carrington indeed has priority for his achievement as being the first to extract usable sunspot data from Thomas Harriot's original manuscripts. While Carrington's priority is somewhat forgotten today (*e.g.*, Vokhmyanin et al., 2020), it is fair to remind the scientific community of his initial achievements in extracting sunspot data from Thomas Harriot's original manuscripts.

Of course, Carrington might have continued his observations and published his results elsewhere. When Carrington's wife died, the *Surrey Advertiser* wrote that: "it is not many weeks since a lengthy article from his [Sc. Carrington's] pen appeared in the Times on 'spots on the surface of the Sun'" (*The Surrey Advertiser*, 1875-11-20, p. 8). The *Surrey Advertiser* allows us to date this "lengthy article" in 1875 and, if confirmed, regard it as Carrington's last article concerning sunspots. There is some ambiguity as to what is meant by "the Times", either *the Surrey Advertiser and County Times*, *the London Times*, or something else, especially since the *Illustrated London News* wrote about Carrington's new observatory. So far, this article has not been uncovered or identified within the existing archival investigations.

## 7. Solar Eclipse as his Possible Motivation

Why, then, did Carrington specifically leave this isolated sunspot drawing on 23 Dec 1870? This is difficult to answer, as Carrington's notes disclose no reason for making the sketch. Interestingly, however, the drawing is dated the day following a solar eclipse that took place on 22 Dec 1870. It is possible that Carrington revived his interest in the Sun specifically for this astronomical spectacle. According to his correspondence (RGO 6/131 of the Cambridge University Archives), Carrington expressed interest in the eclipse.

[9] Wolf (1858, pp. 132-133) explicitly associated his data with Carrington. Wolf stated as follows: "In contrast, Harriot has the undeniable merit of having been the first to carry out a longer series of consistent observations of sunspots, and in this respect his observations are extraordinarily important for the present day, infinitely more important than the any data lacking and purely history related work of Fabricius. I hereby publicly express my warmest thanks to Mr. Carrington for his great effort in making Harriot's work completely available to me. Mr. Carrington has, on the one hand, provided me with a summary table of Harriot's observations, exactly in the form as I have published my own observations thus far, which he has partly derived from Harriot's drawings and partly from his notes regarding spot-free days. On the other hand, he has sent me a copy of the drawings themselves and some samples of the notes. In this current communication I attach below the part of the table which, after reduction to the new style, covers the year 1612, with the monthly relative numbers [NB: sunspot numbers] I have calculated, and I add here the few remaining observations which fall into December 1611 and January 1613, together with the relative numbers [NB: sunspot numbers] derived from them" (Wolf, 1858, pp. 132-133).

We calculated the local eclipse visibility using (i) the methodology of Hayakawa et al. (2021), (ii) the ephemeris data of NASA JPL DE 441 (Park et al., 2021), and (iii) the Earth's rotation parameter of ΔT = 2 s (Stephenson et al., 2016; Morrison et al., 2021), as shown in Figures 8 and 9. On this basis, the totality path passed through the southern side of the Mediterranean Sea, Sicily, Greece, and the Black Sea (Figure 8) and Churt also experienced eclipse visibility up to 0.818 in magnitude at 12:24:19 UT (Figure 9)[10].

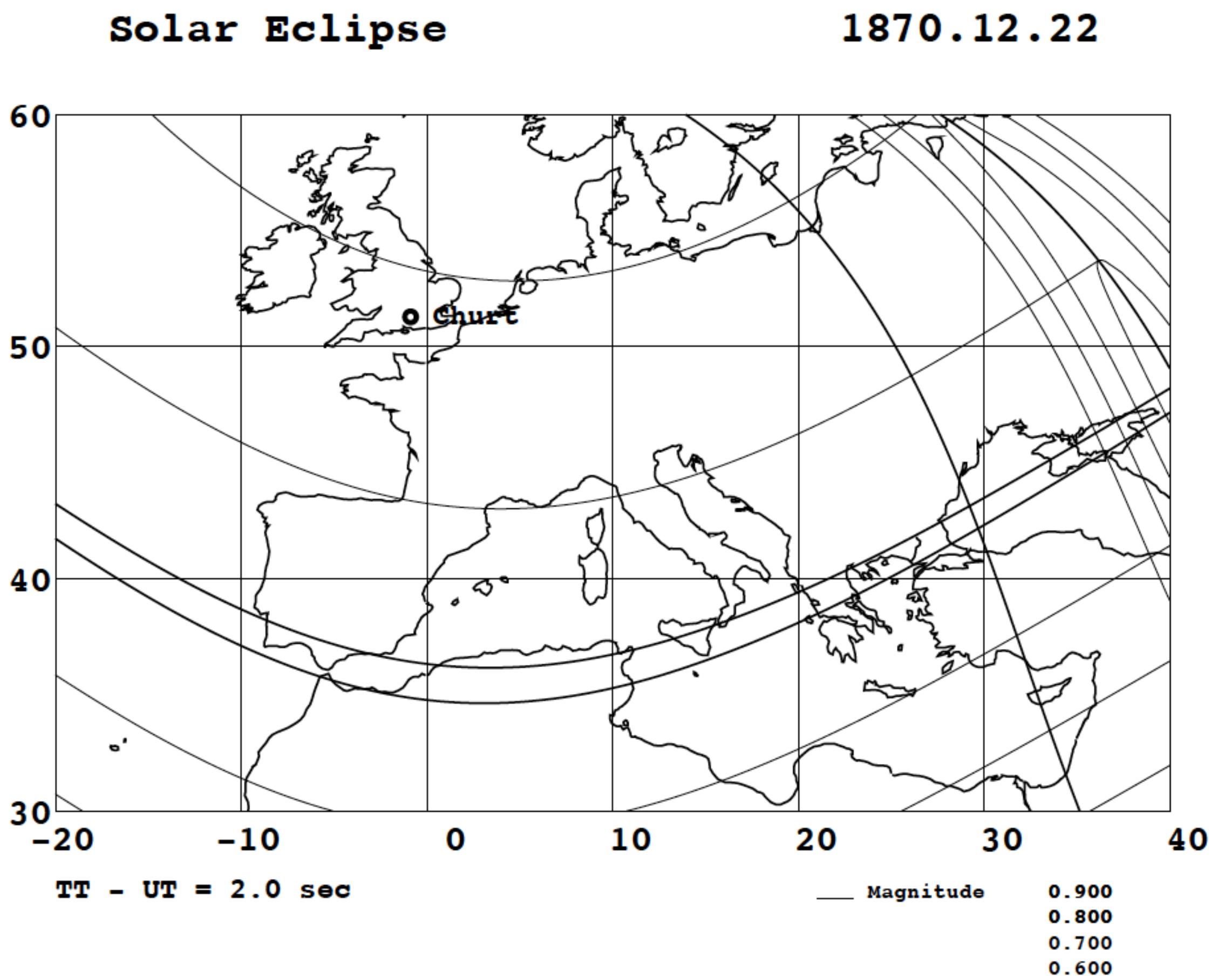

Figure 8: Geographical variation of the local eclipse visibility on 22 Dec 1870, where the thick lines

[10] We obtained these results using (i) the methodology of Hayakawa et al. (2021), (ii) the ephemeris data of NASA JPL DE 441 (Park et al., 2021), and (iii) the Earth's rotation parameter of ΔT = 2 s (Stephenson et al., 2016; Morrison et al., 2021)

indicate the boundaries of totality path and the sunset eclipse visibility and the narrow lines indicate the lines of the eclipse magnitude of 0.9, 0.8, 0.7, and 0.6, respectively.

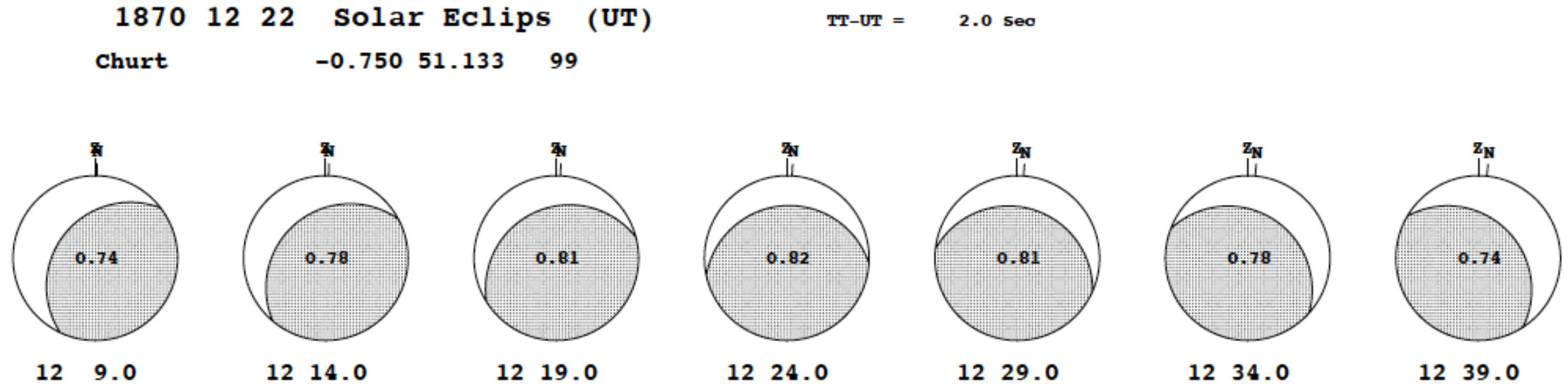

Figure 9: Temporal evolution of the local eclipse visibility at Churt within a range of ±15 minutes the local eclipse maximum (≈ 12:24).

## 8. Summary and Discussions

This study has reported and analysed Carrington's last, unpublished sunspot drawing. The drawing was included in the MSS Carrington 3.3 (f. 277a) of the Royal Astronomical Society. We have dated it at 12:16 LAT on 23 Dec 1870 (Figure 2). The observation was carried out when Carrington was based at Churt Observatory (N51°08'56", W000°45'45"). The drawing allows us to count 3 groups and 10 individual spots on 23 Dec 1870, for which only two data have been identified in previous studies: Schmidt of Athens (G = 8) and Weber of Peckoleh (G = 6), as shown in Figure 6.

Carrington's last sunspot drawing forms a minor addition to the existing datasets of the sunspot group counts (Vaquero et al., 2016; Clette et al., 2023). These additions contribute to the ongoing efforts for the sunspot number recalibration. This conclusion is valid, as long as we follow the standard assumption for the homogeneity of the "Schmidt backbone", as the modern authoritative studies (*e.g.*, Chatzistergos et al., 2017) assume homogeneity on each observer's dataset. However, in reality, Carrington's last drawing is isolated from his Redhill-era observations, was probably made using a different instrument, and shows lower group counts in comparison with contemporaneous observers. In this case, Carrington's last sunspot drawing might instead question the validity of the common assumption of each observer keeping the data homogeneity. This is particularly the case with the "Schmidt Backbone", where Julius Schmidt changed his observational sites and instruments within his data series.

We also need to extend critical reinvestigations on key datasets in the modern authoritative studies

(*e.g.*, Chatzistergos et al., 2017). For example, Koyama changed her observatories, instruments, and scaling factor on some occasions (Figures 6 and 8 of Hayakawa et al. (2020)). Likewise, Cragg lived in California from 1947 to 1976 and then in Australia from 1976 to 2010 (Vaquero et al., 2016, p. 3070). These facts are contrasted with some of the modern authoritative studies (see *e.g.*, Table A.1. and p. 8 of Chatzistergos et al. (2017)), which assume homogeneity on each observer's dataset and used them as reference datasets. Carrington's last sunspot drawing leaves us a lesson on a potential danger on such assumptions and calls for serious investigations on their metadata.

This sunspot drawing allows us to locate these sunspot groups at N14° W37°, N11° W46°, and S08° W08°, where we define the longitude as a longitudinal distance from the central meridian. The groups lie in the low-latitude portion of the sunspot belts as derived from contemporaneous sunspot observations (Figure 7). Carrington's last sunspot drawing additionally forms a minor addition to the existing datasets of the sunspot positions (Muñoz-Jaramillo and Vaquero, 2019; Clette et al., 2023).

Is this a tip of a lost iceberg? That is unlikely. Carrington's last sunspot drawing was dated 23 Dec 1870 in isolation. During the Churt era, Carrington published only one article related to sunspots, where he discussed the question of priority arising from Chevalier's obituaries. This is where Carrington clarified his achievement as being the first (known to the scientific community to date) to extract usable sunspot data from Thomas Harriot's original manuscripts. This priority claim is confirmed from literature surveys on historical investigations into Thomas Harriot's manuscripts. This case brings to mind a caveat. While Wolf's *Mitteilungen* involves a series of historical sunspot datasets before his time, more than a few of their datasets were extracted not by Wolf himself but by other contributors. For example, von Hagen's dataset — one of the backbones of Chatzistergos et al. (2017) — was extracted not by Wolf but by Wagner from the Pulkovo Observatory Archives and contributed to *Mitteilungen*. Unfortunately, the source manuscript is missing. We cannot revisit and recount sunspot groups in von Hagen's original manuscript anymore to check the validity of their claim (Hayakawa et al., 2022, 2026). It is therefore dangerous to assume that all the historical datasets in *Mitteilungen* follow Wolf's sunspot group classifications, while Wolf's own group count methods were also substantially different from modern methods (Svalgaard, 2017).

Otherwise, Carrington may have contributed one 'lengthy article' to 'the Times' on 'spots on the surface of the Sun' (*The Surrey Advertiser*, 1875-11-20, p. 8). This news article has not been located yet. Carrington might have been motivated to direct his telescope to the Sun as a result of the solar eclipse on 22 Dec 1870. At Churt, this eclipse reached as large as 0.818 in magnitude at 12:24:19 UT.

Carrington's manuscripts do not involve any additional sunspot drawings from his Churt era or any explanation as to why he specifically observed the Sun on this date. One might wonder whether this was but one of many sunspot drawings he made during his time at the Churt Observatory. Although not impossible, that is unlikely. Carrington's whole-disk sunspot drawings were bound into MSS Carrington 3.3 and have not left any traces of missing sunspot drawings among his manuscripts. Unfortunately, Carrington's logbooks from the Churt era were left neither in the Royal Astronomical Society Archives nor at Jump House. Further large-scale archival investigations are needed to assess what Carrington observed and intended during his time at Churt.

**Funding**

This research was conducted under the financial support of JSPS Grant-in-Aids JP20H05643 and JP21K13957, the ISEE director's leadership fund for FYs 2021 – 2026, Tokai Pathways to Global Excellence (Nagoya University) of the Strategic Professional Development Program for Young Researchers (MEXT), the young researcher units for the advancement of new and undeveloped fields in Nagoya University Program for Research Enhancement, Transdisciplinary Network linking Space-Earth Environmental Science, History, the Interdisciplinary Research Strategy Projects of the Institute for Space–Earth Environmental Research (ISEE), and Archaeology (JPMXP1324134720) of MEXT Promotion of Development of a Joint Usage/Research System Project: Coalition of Universities for Research Excellence Program (CURE), and the NIHU Multidisciplinary Collaborative Research Projects NINJAL unit "Rediscovery of Citizen Science Culture in the Regions and Today".

**Data Availability**

Carrington's manuscript is preserved in the archives of the Royal Astronomical Society as MSS Carrington 3.3.

**Acknowledgments**

We thank William Tate for kindly accommodating HH's visit to the Jump House and his observatory, Mitsuru Sôma for providing his codes to calculate the heliographic coordinate configurations and the eclipse visibility, Lee Macdonald for his advice on Carrington's instruments, Bruno Besser for his help in translating German text, Sian Prosser for accommodating his visit to the Royal Astronomical Society Archives, and Kate Bond for her marvelous survey and rediscovery of Carrington's photograph.